\documentclass[10pt,conference]{IEEEtran}

\usepackage{cite}

\ifCLASSINFOpdf
   \usepackage[pdftex]{graphicx}
   \graphicspath{{figs/}}
   \DeclareGraphicsExtensions{.pdf,.jpeg,.png}
\else
   \usepackage[dvips]{graphicx}
   \graphicspath{{../figs/}}
   \DeclareGraphicsExtensions{.eps}
\fi
\usepackage[cmex10]{amsmath}
\usepackage{comment}
\usepackage{amsthm}

\usepackage{subfigure}
\usepackage{adjustbox}
\usepackage{framed,multirow}
\usepackage{colortbl}
\usepackage{xcolor}
\definecolor{newcolor}{rgb}{.8,.349,.1}

\usepackage{algorithmic}

\usepackage{array}

\ifCLASSOPTIONcompsoc
  \usepackage[caption=false,font=normalsize,labelfont=sf,textfont=sf]{subfig}
\else
  \usepackage[caption=false,font=footnotesize]{subfig}
\fi
\usepackage{url}

\newif\iffinal
\finaltrue

\iffinal
\else
\usepackage[switch]{lineno}
\fi

\begin{document}

%
\title{Evaluating the Uncanny Valley Effect in Dark Colored Skin Virtual Humans}


\iffinal

\author{\IEEEauthorblockN{Victor Flávio de Andrade Araujo$^*$, Angelo Brandelli Costa$^+$ and Soraia Raupp Musse$^*$}
\IEEEauthorblockA{Graduate Program in Computer Science$^*$, Graduate Program in Psychology$^+$\\
Pontifical Catholic University of Rio Grande do Sul\\
Email: victor.araujo@edu.pucrs.br,\{angelo.costa,soraia.musse\}@pucrs.br}
}

\else
  \author{Sibgrapi paper ID: 30 }
  \linenumbers
\fi

\maketitle

\let\thefootnote\relax\footnote{Published at SIBGRAPI 2023}
\begin{abstract}
With the rapid advancement of technology, the design of virtual humans has led to a very realistic user experience, such as in movies, video games, and simulations. As a result, virtual humans are becoming increasingly similar to real humans. However, following the Uncanny Valley (UV) theory, users tend to feel discomfort when watching entities with anthropomorphic traits that differ from real humans. This phenomenon is related to social identity theory, where the observer looks for something familiar. In Computer Graphics (CG), techniques used to create virtual humans with dark skin tones often rely on approaches initially developed for rendering characters with white skin tones. Furthermore, most CG characters portrayed in various media, including movies and games, predominantly exhibit white skin tones. Consequently, it is pertinent to explore people's perceptions regarding different groups of virtual humans. Thus, this paper aims to examine and evaluate the human perception of CG characters from different media, comparing two types of skin colors. The findings indicate that individuals felt more comfortable and perceived less realism when watching characters with dark colored skin than those with white colored skin. Our central hypothesis is that dark colored characters, rendered with classical developed algorithms, are considered more cartoon than realistic and placed on the left of the Valley in the UV chart.
\end{abstract}


\IEEEpeerreviewmaketitle

\section{Introduction}
\label{sec:intro}
Computer Graphics (CG) has advanced in recent years. One example is the recent technology of MetahumansCreator\footnote{https://www.unrealengine.com/en-US/metahuman}, which aims to model and animate virtual humans with high levels of realism. 
According to the Uncanny Valley theory~\cite{mori2012uncanny}, concerning user experience, it is essential to create virtual humans similar to human beings. In general, few methodologies consider human diversity for studies of human perception~\cite{zell2019perception}, such as gender and skin color, when studying UV theory. One example is the work of Araujo et al.~\cite{araujo2021analysis}, which states that women tend to feel more comfortable than men when watching realistic female CG characters. 
Human perception should be the main focus of virtual human design for both industry and academic research. In particular, creating CG characters for different media should involve human representation that reaches all groups. For example, in work by Saneyoshi et al.~\cite{saneyoshi2022other}, the authors evaluated the other-race in the UV effect concerning realism variations in a continuum of two virtual humans, one Japanese, and the other European. The authors concluded that in terms of perceived pleasantness, in-group participants did not tolerate slight differences. According to Brown~\cite{brown2020social}, Social Identity theory explains that people prefer to see themselves and their groups in a positive light, which implies that they look for positive distinction in their perceptions and relationships over other groups. Both in the field of psychology and in computing, there are studies on in-group advantages~\cite{elfenbein2002there, krumhuber2015real}
, that is, people in a group recognize and encode better the facial characteristics of real or virtual humans from their same group. 
In addition, there are few studies with results focused on out-group advantages~\cite{saneyoshi2022other, hasler2017virtual} concerning virtual humans and UV, that is, people from a minority group evaluating virtual humans from a different group in a similar way to people belonging to the same group.

According to Kim et al.~\cite{kim2021countering}\footnote{www.youtube.com/watch?v=ROuE8xYLpX8}\footnote{https://yaledailynews.com/blog/2022/03/09/yale-professors-confront-racial-bias-in-computer-graphics/}, multiple dimensions of diversity should be considered. The authors stated that most studies on CG from the scientific community present results based on virtual humans with white skin colors. Specifically, the authors mention that most techniques for creating skin color through CG are based on white skin color and that the emphasis on lighting is exaggerated when applied to dark skin. In this case, a question arises, do dark colored skins convey to people a lower level of realism than white colored skins? If this is true, following the original UV theory~\cite{mori2012uncanny} and other studies that support it (\cite{macdorman2016reducing, macdorman2009too}), reducing realism in virtual humans may change the perceived eeriness/discomfort. 
Bringing it to the context of our work, can characters with dark colored skin cause low or high sensations of discomfort in human perception?

The goal of this paper is to investigate people's perceptions about dark colored skin CG characters already established by various media to try to answer the questions that were raised during the last paragraphs. Looking at some studies that evaluated the effect of UV on CG characters from various media, dark colored skin characters seem to be perceived with higher comfort values~\cite{flach2012evaluation, araujo2021perceived}, if compared to characters with white colored skin. Therefore, 
we raise the following hypotheses to be studied in our work:

\begin{itemize}
    \item $H0_1$ defining that the UV effect is similar for white and dark colored skin characters, with similar levels of realism. First, to test this hypothesis, we used the data of white colored skin characters provided by the work of  
    Araujo et al.~\cite{araujo2021perceived}. To compare these results, we recreated the experiment proposed by Araujo et al. only with dark colored skin characters. 
    \item $H0_2$ defines that UV's effect on different colored skin characters is similar for people with different racial identifications. 
    By testing this hypothesis, we can discuss in-group and out-group advantages. 
\end{itemize}

The paper contributes to the UV theory concerning virtual characters with dark colored skin. By testing $H0_1$, we investigate the diverse perceptions of realism and comfort when comparing characters with varying skin colors. Additionally, in considering $H0_2$, we can examine the impact of in-group and out-group advantages in the context of the UV.

\section{Related Work}
\label{sec:relatedWork}


Some studies can help us to understand in-group and out-group advantages. In the work of Beaupre and Hess~\cite{beaupre2005cross}, the authors investigated cultural differences in facial recognition accuracy and in-group advantage for emotion recognition
. 
The author mentioned that individuals from the same group could judge emotional expressions more easily due to factors such as familiarity with facial morphology.
In the same line of morphological questions, Balas and Nelson~\cite{balas2010role} evaluated facial shape and skin pigmentation 
information in virtual humans through electrophysiological and behavioral measures of participants. 
In one of the results, participants reported that they did not have many exposures to faces with dark skin. This aspect of in-group advantage over own race is a natural effect of human beings.
In the work of Kang and Lau~\cite{kang2013revisiting}, the authors conducted studies to explore out-group advantages
. In one of the results, one of the groups was European Americans, and the other group was Asian Americans, and both groups were asked to identify imitated emotions and spontaneous emotions from each of the two groups, that is, in-group and out-group. One of the results revealed that an in-group advantage was observed in the spontaneous expressions scenario but not in the mimicked expression condition. In work by Hehman et al.~\cite{hehman2010division}, the authors investigated whether the out-group question would result from social categorization and not perceptual experience. 
Results showed that participants had related memory for own-race faces when grouped by race/ethnicity. When grouped by the university, participants had positive results for the faces of the university itself. 


Several studies regarding virtual humans' perception correlate with race/ethnicity and skin color. 
The work of Bedder et al.~\cite{bedder2019mechanistic} explains implicit bias towards the negative evaluation of people in the out-group when using a virtual body
. In another work by Banakou et al.~\cite{banakou2016virtual}, the authors mention that incorporating white participants into a black virtual body is associated with an immediate decrease in their implicit racial prejudice against black people. 
In one of the results, the authors showed that the implicit bias decreased more for those with a black virtual body than for white ones. 
In work by Hasler et al.~\cite{hasler2017virtual}, the results showed that white participants who used the virtual dark skin color expressed greater mimicry when they saw virtual groups of dark skin colors. The authors mentioned that these white participants treated groups of black virtual humans from the environment as their in-group, while the white virtual humans were treated as an out-group. 

In all these studies, we saw the importance of cultural exposure and diversity for people's perception of virtual characters/humans with dark colored skin. As mentioned in the course presented by Zell et al.~\cite{zell2019perception}, studying the perception of virtual characters is essential for the design of these characters. Therefore, the virtual human's design must be considered for the user experience. When we talk about the concepts of in-group and/or out-group advantages~\cite{saneyoshi2022other, hasler2017virtual, elfenbein2002there, krumhuber2015real}, it is related to the user experience and representativeness. As mentioned in the work by Katsyri et al.~\cite{katsyri2015review}, humans observe virtual humans and seek human likeness in these artificial beings. Therefore, all in-group and out-group studies are essential to understanding this relationship, and our work tries to follow this line.

\section{Methodology}
\label{sec:methodology}

In our work, we used the dataset provided by Araujo et al.~\cite{araujo2021perceived}, which contains perceptual data about CG characters from different media. Therefore, we used only the perceptual data of white colored skin characters, i.e., we did not present the white skin color characters to the participants in our work. 
 In addition, we evaluated the participants' perception 
 with respect to the dark colored skin characters. 
 Therefore, this section is organized as follows: in Section~\ref{sec:characters}, we present the studied characters, 
and in Section~\ref{sec:questions_and_stimuli}, we present the questions and the stimuli presented to the participants.

\subsection{The Characters}
\label{sec:characters}

\begin{figure*}[!htb]
  \centering
  \subfigure[icon][Movie: \newline The Incredibles 2]{\includegraphics[ width=0.11\linewidth, height=0.11\linewidth]{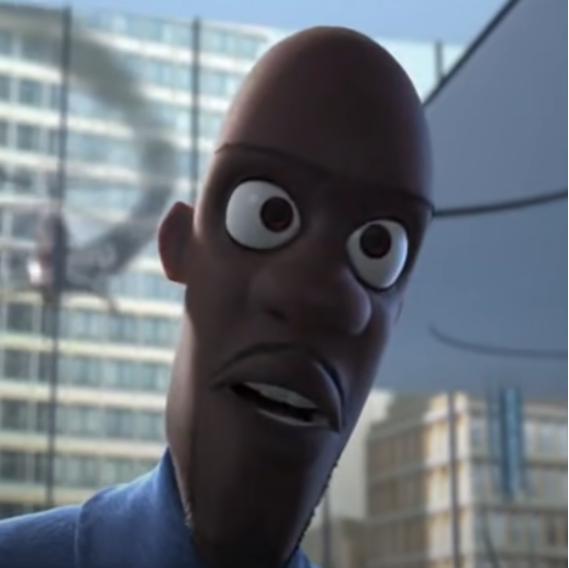}}
  \subfigure[icon][Movie: Soul]{\includegraphics[width=0.11\linewidth, height=0.11\linewidth]{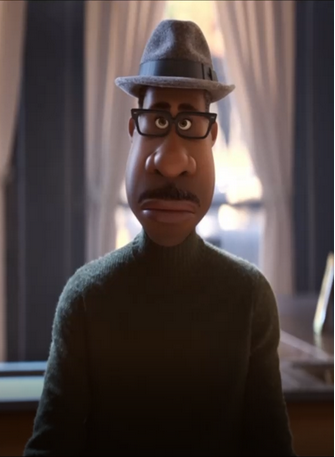}}
  \subfigure[icon][Series: Arcane]{\includegraphics[width=0.11\linewidth, height=0.11\linewidth]{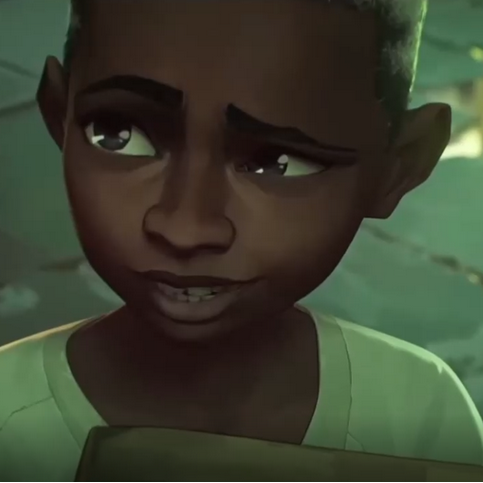}}
  \subfigure[icon][Game: GTA San \newline  Andreas]{\includegraphics[width=0.11\linewidth, height=0.11\linewidth]{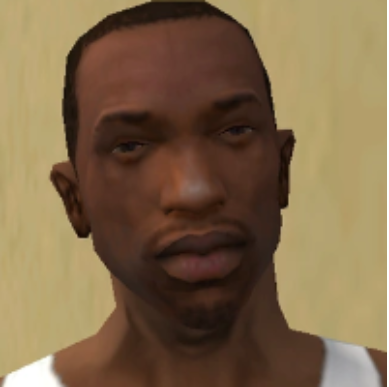}}
  \subfigure[icon][Game: The \newline Walking Dead \newline from Telltale]{\includegraphics[width=0.11\linewidth, height=0.11\linewidth]{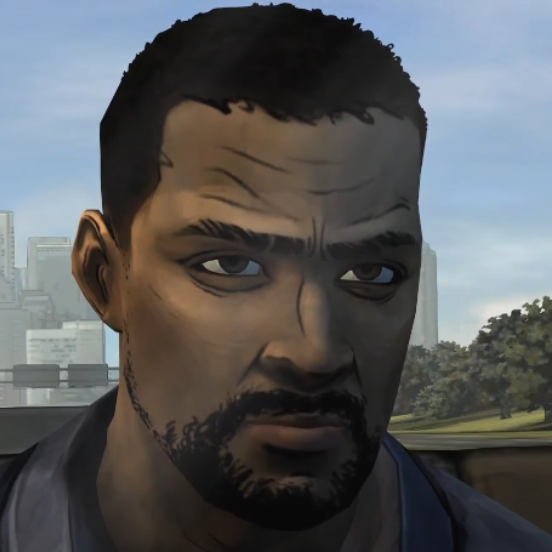}}
  \subfigure[icon][Movie: Encanto]{\includegraphics[width=0.11\linewidth, height=0.11\linewidth]{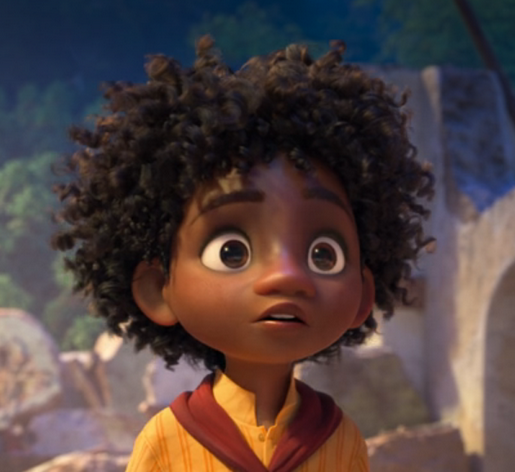}}
  \subfigure[icon][Movie: \newline Spider-Verse]{\includegraphics[width=0.11\linewidth, height=0.11\linewidth]{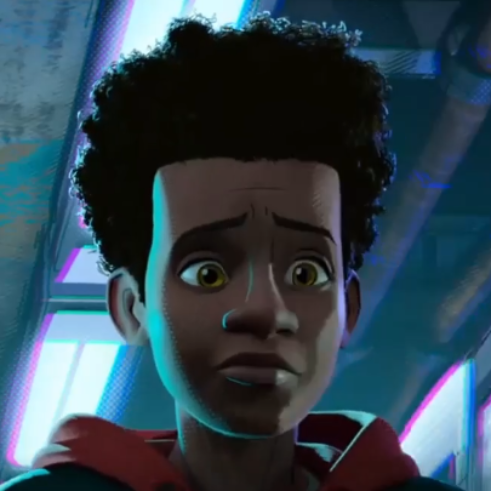}}
  \subfigure[icon][Movie: Moana]{\includegraphics[width=0.11\linewidth, height=0.11\linewidth]{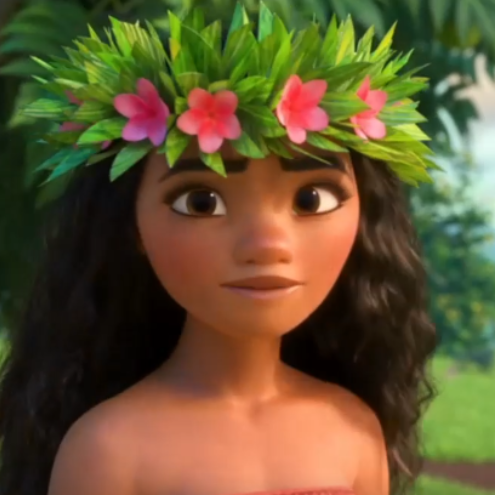}}\\
  
  \subfigure[icon][Game: \newline True Crime \newline New York City]{\includegraphics[width=0.11\linewidth, height=0.11\linewidth]{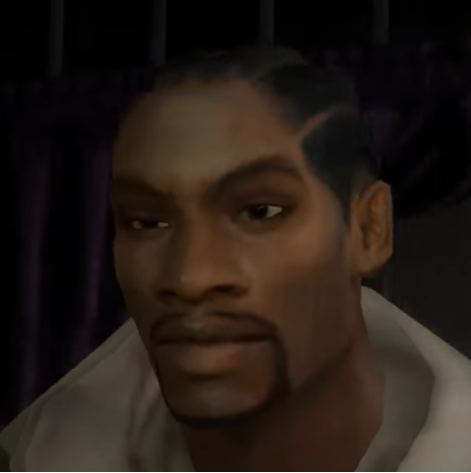}}
  \subfigure[icon][Internet: \newline Obama's Cartoon]{\includegraphics[width=0.11\linewidth, height=0.11\linewidth]{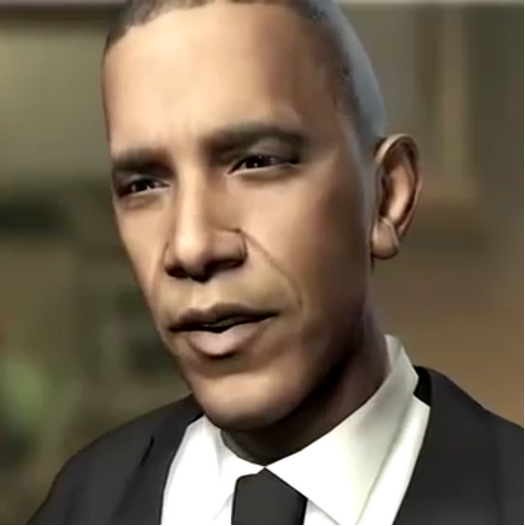}}
  \subfigure[icon][Game: GTA V]{\includegraphics[width=0.11\linewidth, height=0.11\linewidth]{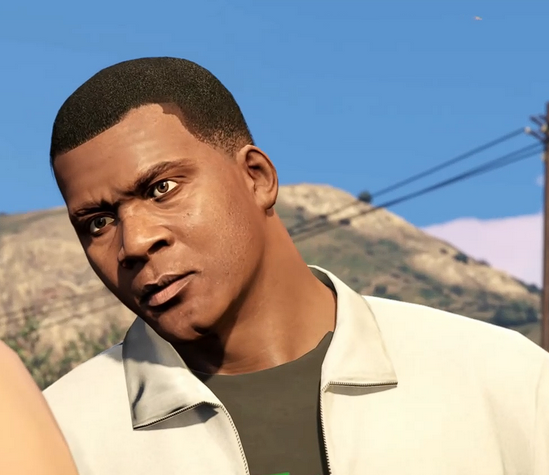}}
  \subfigure[icon][Game: \newline Mortal Kombat 11]{\includegraphics[width=0.11\linewidth, height=0.11\linewidth]{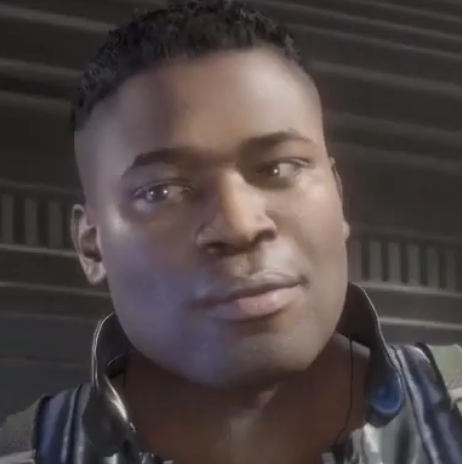}}
  \subfigure[icon][Game: Fifa 19]{\includegraphics[width=0.11\linewidth, height=0.11\linewidth]{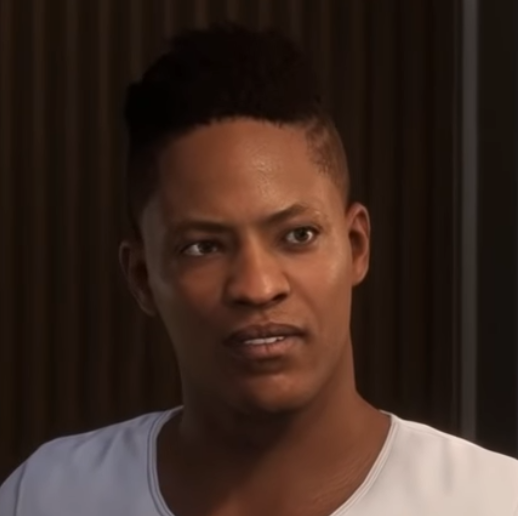}}
  \subfigure[icon][Game: \newline Call of Duty \newline Black Ops 2]{\includegraphics[width=0.11\linewidth, height=0.11\linewidth]{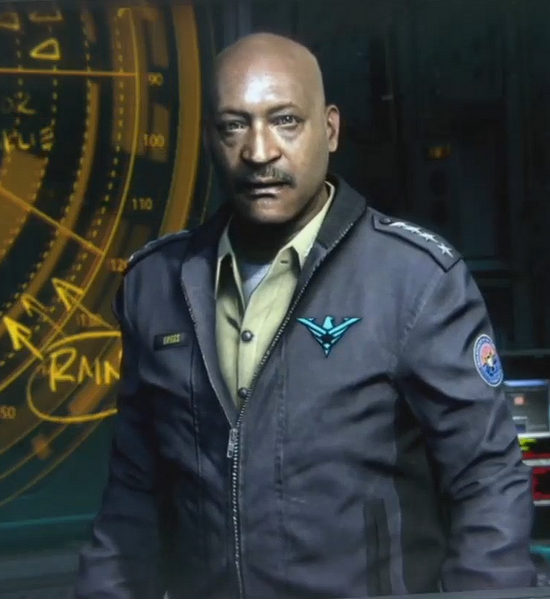}}
  \subfigure[icon][Game: \newline Horizon Zero  \newline Down]{\includegraphics[width=0.11\linewidth, height=0.11\linewidth]{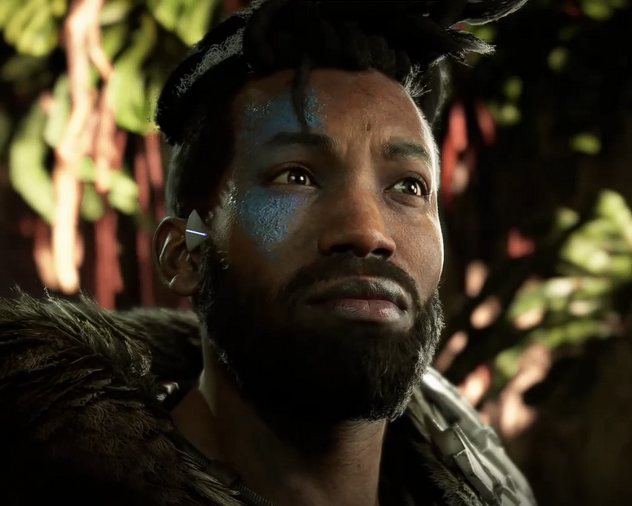}}
  \subfigure[icon][Internet: \newline MetaHuman Creator]{\includegraphics[width=0.11\linewidth, height=0.11\linewidth]{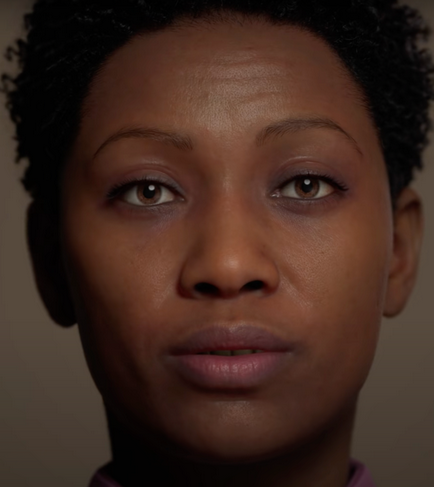}}
  
  \caption{All dark colored skin characters presented in this work.}
  \label{fig:black_characters_figures}
\end{figure*}


To test $H0_1$ ("defining that the UV effect is similar for white and dark colored skin characters, with similar levels of realism"), we used perceptual data from white colored skin characters from the dataset by Araujo et al.~\cite{araujo2021perceived}. Remember that, concerning skin color and race/ethnicity, we focus only on visual attributes; that is, we do not use narratives and contexts involving the characters in the media. Therefore, we did not use data from six of the 22 characters in the dataset. 
Indeed, we removed characters (analyzed data) that, in our view, did not have white skin color and characters that we were not sure about their skin color, e.g., 
Hulk and his green skin color. 
Considering the UV theory, these characters removed from the analysis do not negatively influence the methodology for creating the comfort chart (based on the UV chart), because the three levels of realism proposed in the work of~\cite{katsyri2015review} still remain: Unrealistic, Moderately Realistic, and Very Realistic (the characters with white colored skin can be seen in Figure~\ref{fig:comfortChartQ5}(b)). 
These classifications were given by responses from participants in the previous work (\cite{araujo2021perceived}. 

Regarding the dark colored skin characters chosen for our experiment, we 
chosen characters to be classified in the three levels of realism, as in the group of white colored ones
, as discussed by Katsyri et al.~\cite{katsyri2015review}. 
Figure~\ref{fig:black_characters_figures} shows the characters used in our work. 
Following the methodology proposed in previous work~\cite{araujo2021perceived}, we looked for unrealistic characters (they are usually cartoon characters, 
for example, characters "a", "b", "f", "g" and "h"
), moderately realistic characters (they are not a cartoon, but they do not have high levels of realism, for example, "c", "d", "e", "i", "j" and "l"), and realistic characters ("k", m", "n", "o" and "p"). We believe that stimuli involving different characters from different media can increase the subjectivity of evaluations, which can be a limitation of our work. However, we needed to follow the methodology presented in the previous work to reduce differences in comparisons between perceptual data related to dark and white colored skin characters.
As these characters are "public ready", 
we looked for the characters in Youtube videos, channels with many followers, and videos with many views that had permission for content transmission
\footnote{Still, we invoke the "fair use policy: Copyrighted images reproduced under ``fair use policy'' for protected images and videos.}.

\subsection{Questions and Stimuli}
\label{sec:questions_and_stimuli}

Before asking questions related to the characters, we asked participants the following demographic questions: \textit{i)} The first question was "In terms of skin color and race/ethnicity, do you identify as a person:" with the answer options "Black", "Indigenous", " Yellow", "Brown", "White", "Other" (with a free text field), and "I do not want to answer"\footnote{Those are the official categories of (removed for the blind review) geographic institute used in the census}. Note: As the previous work did not have the question of color and race/ethnicity, 
we did not use such data 
in the analyzes of $H0_2$; 
\textit{ii)} "Do you identify with the gender:" with the options "Female", "Male", "Nonbinary", "Other" (with a free text field), and "I do not want to answer";
\textit{iii)} "Age:" with the options "18 to 20 years old", "21 to 29", "30 to 39", "40 to 49", "50 to 59" and "Over 60"; \textit{iv)} "Inform the level of education" with the options "Incomplete high school", "Complete high school", "Complete higher education", and "Postgraduate".

In the previous work~\cite{araujo2021perceived}, the authors asked participants to answer five questions about the characters, for each image or video. However, as we want to focus only on perceived realism and comfort (a possible UV chart), we removed the questions that were not the focus of our work. With this, we used only two questions, as the following: 
\textit{i)} Question Q1 (''How realistic is the character") which the possible answers were ''Unrealistic", ''Moderately realistic", and ''Very realistic", 
is a 3-Likert Scale. 
Q1 is responsible for shaping the order of characters in terms of Realism, that is, each character has a response average between 1-Unrealistic and 3-Very Realistic, and we rank them based on these averages; 
and \textit{ii)} Q2 (''Do you feel some discomfort/strangeness looking at this character?"), which was a ''Yes" or ''No" question. We used the percentage of "No" answers for each character, that is, how much comfort was perceived. 

We divided the characters into two stimulus stages. 
In the first stage, we presented one 
image of each character, while in the second stage, we presented one video. All media were presented in a random order to avoid bias. 
The images, videos, and questions were applied through an online form created with Google Forms and transmitted to volunteer respondents through social networks (Whatsapp and Instagram groups). 
We presented a 
term of commitment approved by the ethics committee\footnote{Research Ethics Committee of Pontifical Catholic University of Rio Grande do Sul, Brazil - Project Number: $46571721.6.0000.5336$}, where 
we presented all the possible risks of the respondent's participation. 
The images and videos of the characters have the same size in the form, approximately 520x520, measured on a 15-inch Full HD 1080 screen. 
The videos had an average duration of $4.937$ seconds and it was focused on the characters' faces in both images and videos to avoid body bias.

\section{Results}
\label{sec:results}

The questionnaire was 
answered by 82 volunteers, from which: 81.7\% answered "White", 13.4\% answered "Brown", and 4.9\% "Black" for the question of identification with skin color and race/ethnicity. In this case, as we had many more white participants, we separated the data into white and brown/black (in Brazil, black and brown people can be grouped into one category~\cite{gomes2019ser}) participants as they declared themselves. In addition, participants were 54.9\% women and 41.5\% men; 41.1\% were between 18 and 20 years old; 48.8\% had completed high school at most
. In all statistical analyses, we used 5\% of significance level (\textit{Factorial ANOVA} 
test).



This section evaluates the $H0_1$ ("defining that the UV effect is similar for white and dark colored skin characters, with similar levels of realism") and the $H0_2$ ("defining that the effect of UV on different colored skin characters is similar for people with different racial identifications.") hypotheses. So, we compared white colored skin characters (Araujo et al's work dataset
) and dark colored skin characters (our work, Figure~\ref{fig:black_characters_figures}) in a UV theory perspective, that is, perceptions of realism 
and comfort. 

\subsection{Realism}
\label{sec:human_likeness}

\begin{figure*}[!htb]
  \centering
  \subfigure[fig:blackorder][Dark Colored Skin Characters]{\includegraphics[width=15cm]{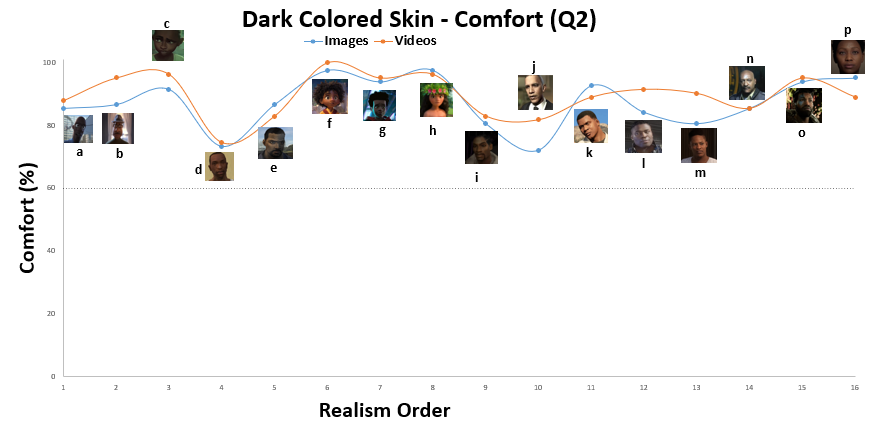}}
  \subfigure[fig:whiteorder][White Colored Skin Characters]{\includegraphics[width=15cm]{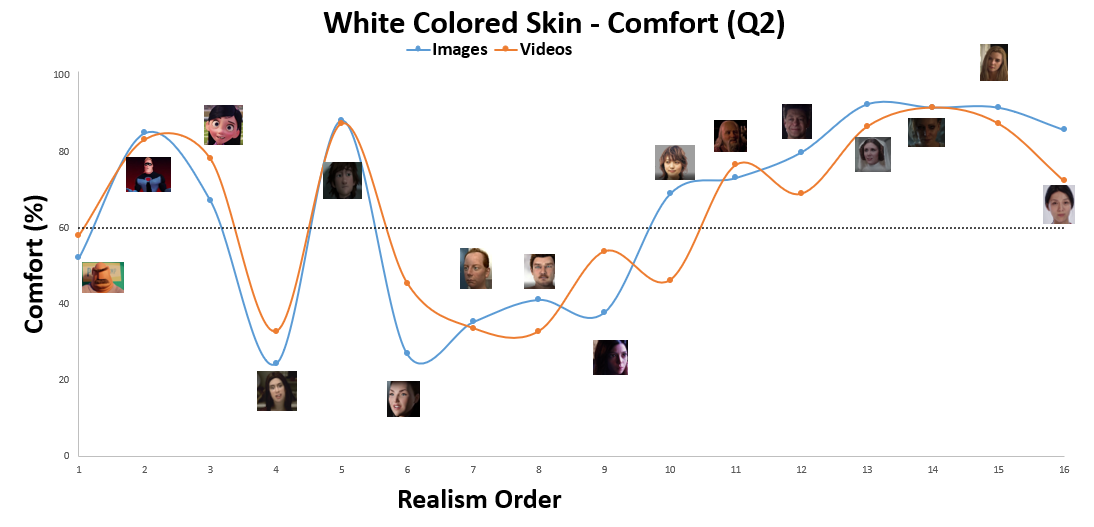}}
    \caption{
    In a), There are all dark colored skin characters evaluated in this work. The characters were ordered according to the perceived Realism (Q1 question). The blue and orange lines represent the comfort percentages (Q2 question) regarding images and videos, respectively. The dashed line represents 60\% of comfort values. In b) all white colored skin characters from~\cite{araujo2021perceived} are illustrated in this work.
    }
    \label{fig:comfortChartQ5}
\end{figure*}

In Figure~\ref{fig:comfortChartQ5}(a), we can see 
the ordering of the dark colored skin characters in our work on Realism (horizontal axis)
. 
To order the characters, 
we used each character's averages of realism (answers to question Q1) 
It is essential to notice that the analyzed Q1 question only used the answers regarding the pictures, while Q2 analyses also used the perceptual answers regarding the presented videos. 
Regarding $H0_1$, 
we found a significant result in the main effect of Skin Colored feature (F(1, 3214)=$22.96$  \textit{p}$<.001$), in which the average realism of white colored skin characters 
was $2.049$, and in ours, the average of dark colored skin characters was $1.9$. With this, \textbf{we can say that dark colored skin characters were considered less realistic than the white colored skin characters, so this refutes partially $H0_1$ 
hypothesis because participants do not evaluate the realism of black and white colored characters similarly.} The second part of the hypothesis, focused on the perception of comfort, will be presented in Section~\ref{sec:comfort_and_uvEffect}. 

Regarding $H0_2$, 
we did not find a significant main effect of the Participant's Race
. Remembering that, as in the previous work 
we did not have racial identification data, we used only our data in such analysis. That is, participants who identified themselves as white versus participants who identified themselves as brown/black in relation to dark colored skin characters (respectively, averages of $1.89$ and $1.91$). 
In this case, with this, \textbf{we cannot refute the $H0_2$, in terms of perceived realism, for two reasons: First, we do not know the racial identification of the participants in the baseline study, and second, we did not find a significant result between the different racial groups present in our work.}

\subsection{Comfort and UV effect}
\label{sec:comfort_and_uvEffect}

Concerning the perceived comfort (Q2 question), the Figure~\ref{fig:comfortChartQ5} also shows the perception of comfort (Y axes)
. Contrary to what happens with white colored characters (illustrated in Figure~\ref{fig:comfortChartQ5}(b)), we can see that the characteristic shape of UV theory does not appear in our work (Figure~\ref{fig:comfortChartQ5}(a)), i.e., the visual aspect of the Valley. 
Considering the comfort values, in (a), all dark colored skin characters were above 60\%, whereas in (b), six white colored skin characters had comfort values below 60\% for videos, and five for images. 
Concerning 
$H0_1$, 
both for image and video (respectively, F(1, 3214)=$212.64$  and \textit{p}$<.001$, and F(1, 3214)=$277.58$  and \textit{p}$<.001$), we found significant results in the main effects of Skin Colored feature. The white skin colored characters had average comfort percentages of 65.07\% in the image and 64.65\% in the video, while the dark colored skin characters in our work had 87.27\% and 89.55\%. Therefore, \textbf{we can say that dark colored skin characters conveyed more comfort than white colored skin characters, which fully refutes hypothesis $H0_1$ (the evaluation of realism was presented in Section~\ref{sec:human_likeness}). 
Looking the Figure~\ref{fig:comfortChartQ5}, 
we can also say that the UV effect was different for dark colored and white colored skin characters, refuting the $H0_1$ hypothesis.}

Regarding $H0_2$, as well as in the realism analysis, we did not find significant results. 
As in the analysis of realism, the values of perceived comfort about dark colored skin characters were higher for black/brown participants (respectively, 89.73\% and 90.62\%) than for white participants (86.66\% and 89.17\%). 
With this, \textbf{we can say that the perception of comfort about dark colored skin characters was similar for participants with different racial identifications, that is, confirming the $H0_2$ hypothesis. 
However, we need to point out that this could change if the sample of the black/brown participants were larger.}


\section{Discussion}
\label{sec:discussion}

In this section, we discuss some results presented in Section~\ref{sec:results}. 
Firstly, regarding the analyzes involving $H0_1$, we compared the perception of all participants in the work from Araujo et al.~\cite{araujo2021perceived} about white colored skin characters and the perception of all participants in our work about dark colored skin characters. 
The results showed us some interesting points: the first is that the UV effect~\cite{mori2012uncanny} was different for the dark colored skin characters, which conveyed more comfort and were considered less realistic than the white colored skin characters. These results answer our question in Section~\ref{sec:intro} "Do characters with dark colored skin cause low or high sensations of discomfort in human perception?". As a consequence, in terms of UV theory, 
the Figure~\ref{fig:comfortChartQ5}(a) did not present the expected Valley, i.e., not presenting characters with lower values of comfort. 
We can hypothesize at least two situations that could interfere with human perception causing the high values of comfort obtained in this work. Firstly, this may be related to rendering and illumination techniques. 
Kim~\cite{kim2021countering} mentioned that most CG techniques are focused on virtual humans with white colored skin and that dark colored skin virtual humans often have exaggerated skin color lighting. 
Still, according to Kim, techniques such as "translucency and the corresponding physical mechanism of subsurface scattering has become synonymous with human skin in rendering. However, translucency is only the dominant visual feature of young, white Europeans and fair-skinned East Asians". In this case, the author still cited several scientific studies on this subject, which involve only examples of virtual humans with white skin color, which becomes a scientific bias in creating future virtual humans. In the same line, the observed decrease in perceived realism of dark colored characters may be related to the theory of Dehumanization~\cite{kteily2022dehumanization}, since "algorithmically" the main methods were modeled and tested with white colored skin~\cite{kim2021countering}. So, we can hypothesize that, in their perceptions, the participants did not "humanize" the dark colored skin characters, which could indicate an issue with rendering methods. Therefore, this question needs further study to draw more substantial conclusions. 
While one explanation may be the rendering and illumination techniques, another possibility can be the extension of human in-group and out-group effects. However, as we did not have many participants who considered themselves black/brown, we did not achieve significant results in this aspect. 
Lastly, it is important to mention why this type of research is relevant for the CG community. 
According to Katsyri et al.~\cite{katsyri2015review} and Araujo et al.~\cite{araujo2021analysis}, the experience we have when we observe an artificial being is related to human identification. So, our ultimate goal is to contribute with methodologies that provide more equity and representation in the modeling of virtual humans, providing better experiences to users.

\section{Final Considerations}
\label{sec:finalConsiderations}

This study investigated the perception of dark-colored skin computer-generated (CG) characters that have been prominently featured in various forms of media, such as movies, games, and series. We employed the Araujo et al.~\cite{araujo2021perceived} dataset to compare the perceptual data associated with characters with white skin color with our own collected data. Our dataset is specifically concentrated on dark-colored skin CG characters, excluding those with lighter skin tones. Additionally, we explored the UV effect on these perceptual judgments. As a result, our contributions extend to the realms of UV theory and the perception of virtual humans.

Concerning our initial hypothesis ($H0_1$ - "defining that the UV effect is similar for white and dark colored skin characters, with similar levels of realism"), our findings unveiled a notable disparity. The results indicated that the UV effect varied distinctly between the perceptual data about dark-colored skin characters and that of white-colored skin characters. Additionally, participants in our study reported feeling more comfortable and perceiving lower levels of realism when engaging with dark-colored skin characters, in contrast to those participants in the previous study, who expressed their perceptions of white-colored skin characters. 

Regarding our second hypothesis ($H0_2$ - "defining that the effect of UV on different colored skin characters is similar for people with different racial identifications"), 
our findings shed interesting light. We discovered that most participants who self-identified as black/brown exhibited perceptual responses similar to those who identified as white when evaluating dark-colored skin characters. However, two notes are essential in this context. Firstly, we could not compare with the perception of white colored characters because the baseline work did not have information about the participants. Secondly, the number of participants identifying as black/brown was relatively small, warranting further investigation in future studies to obtain a more comprehensive understanding of this phenomenon.

Lastly, we would like to emphasize the significance of this research area within the CG community, as it holds the potential to enhance and ensure a more equitable and inclusive user experience. By delving into the perception of dark colored skin characters and investigating the UV, we aim to contribute towards creating a more comprehensive understanding of virtual human representation. This knowledge can guide the development of CG content that is visually appealing and promotes fairness, diversity, and a greater sense of belonging for all users. 


\section*{Acknowledgment}

The authors would like to thank CNPq and CAPES for partially funding this work.



\bibliographystyle{IEEEtran}
\bibliography{example}
%
%


\end{document}